\documentclass[12pt,sort,sort&compress]{revtex4}

\usepackage{amsmath,amsfonts,amssymb,amsthm}
\usepackage{xy}
    \xyoption{matrix}
    \xyoption{arrow}

\def\mathscr{\mathcal}

\def\be{\begin{equation}}
\def\ee{\end{equation}}
\def\bel#1{\begin{equation} \label{#1}}
\def\beq{\begin{equation}}
\def\eeq{\end{equation}}

\def\nn{\nonumber}
\def\bea{\begin{eqnarray}}
\def\eea{\end{eqnarray}}
\def\lab{\label}

\def\To{\longrightarrow}
\def\V{\mathcal{V}}
\def\M{\mathcal{M}}
\def\H{\mathcal{H}}
\def\B{\mathcal{B}}

\def\W{\mathcal{W}}
\def\A{\mathcal{A}}
\def\T{\mathcal{T}}
\def\U{\mathcal{U}}
\def\F{\mathcal{F}}

\def\C{\mathbb{C}}

\def\R{\mathbb{R}}
\def\N{\mathbb{N}}

\def\d{\partial}

\def\eps{\varepsilon}
\def\vp{\varphi}
\def\up{\uparrow}

\def\Tr{\operatorname{Tr}}

\def\Aut{\operatorname{Aut}}
\def\tr{\operatorname{tr}}

\def\Lie{{\operatorname{Lie}}}

\def\<{\langle}
\def\>{\rangle}
\def\half{\frac{1}{2}}

\newtheorem{lemma}{Lemma}
\newtheorem{theorem}{Theorem}

\newtheorem{definition}{Definition}
\newtheorem{axiom}{Axiom}
\theoremstyle{remark}
\newtheorem{remark}{Remark}

\def\tilde#1{\widetilde{#1}}

\def\b#1{\mathbf{#1}}

\def\te#1{\text{#1}}
\def\bar#1{{\overline{#1}}}

\def\f#1#2{\frac{#1}{#2}}

\def\dom{\operatorname{dom}}

\def\s{\mathscr{S}}
\def\S{\underline{\s}}
\def\ol{\overline}
\def\I{\mathbf{1}}
\def\rad{\operatorname{rad}}
\def\Op{\operatorname{Op}}
\newcommand{\ox}{\otimes}
\newcommand{\os}{\otimes_s}

\def\eqref#1{(\ref{#1})}
\newcommand{\er}{\eqref}

\newcommand{\gl}{\operatorname{gl}}

\def\Hphys{\mathcal{H}_{\text{phys}}}

\def\A{\mathcal{A}}
\def\uA{\underline{\A}}

\newcommand{\g}{\mathfrak{g}}

\newcommand{\ST}{\Sigma_T}

\def\al{\alpha}
\newcommand{\stalg}{$\ast$-algebra~}

\newcommand{\til}{\tilde}
\newcommand{\w}{\omega}
\def\frakw{\mathfrak{w}}
\def\m{\mu}
\def\n{\nu}

\newcommand{\SymmetryGp}{{\cal G}}
\renewcommand{\O}{{\cal O}}
\newcommand{\Mat}{\mathrm{Mat}}
\newcommand{\Om}{\Omega}
\renewcommand{\b}{\beta}
\newcommand{\la}{\lambda}
\newcommand{\ve}{\varepsilon}

\def\wt{W_{\rm tree}}

\begin{document}

\title{A General Theory of Wightman Functions}

\author{William Gordon \surname{Ritter}}
\affiliation{{\small\textit Harvard University Department of Physics} \\
    {\small\textit 17 Oxford St., Cambridge, MA 02138}}

\date{\today}

\begin{abstract}
One of the main open problems of mathematical physics is to
consistently quantize Yang-Mills gauge theory. If such a
consistent quantization were to exist, it is reasonable to expect
a ``Wightman reconstruction theorem,'' by which a Hilbert space
and quantum field operators are recovered from $n$-point
functions. However, the original version of the Wightman theorem
is not equipped to deal with gauge fields or fields taking values
in a noncommutative space. This paper explores a generalization of
the Wightman construction which allows the fundamental fields to
take values in an arbitrary topological $\ast$-algebra. In
particular, the construction applies to fields valued in a Lie
algebra representation, of the type required by Yang-Mills theory.
This appears to be the correct framework for a generalized
reconstruction theorem amenable to modern quantum theories such as
gauge theories and matrix models. We obtain the interesting result
that a large class of quantum theories are expected to arise as
limits of matrix models, which may be related to the well-known
conjecture of Kazakov. Further, by considering deformations of the
associative algebra structure in the noncommutative target space,
we define certain one-parameter families of quantum field theories
and conjecture a relationship with deformation quantization.
\end{abstract}

\keywords{}


\maketitle


\section{Introduction}

The Wightman axioms were formulated by G\aa rding and Wightman in
the early 1950's, but no nontrivial examples existed at that time,
and consequently the axioms were not published until 1964
\cite{Axioms}, at which time their publication had been motivated
by the Haag-Ruelle scattering theory. The axioms are thoroughly
discussed and many consequences are derived in the two excellent
books \cite{Streater:vi} and \cite{Jost}. We will also formulate
the axioms below in Section \ref{sec:wightax} by way of
introduction.

It is known that the Wightman axioms, in their original and
unmodified form, describe only a small subset of the mathematical
models used in elementary particle physics. Thus, many authors
have considered modifications of the axioms which allow newer and
more exotic physical theories to be formulated as rigorous
mathematics. If we wish to perturb the axioms slightly, one
obvious change with clear-cut physical implications is to relax
the requirement that the test function space be ${\cal S}(\R^4)$.
A large class of alternative test function spaces which still
allow a formulation of the microscopic causality condition were
proposed and developed by Jaffe \cite{Jaffe:1967nb}. The results
of the present paper are a more radical modification, in which the
test functions in ${\cal S}(\R^4)$ are replaced by functions into
a noncommuting $\ast$-algebra.

There are at least two types of equivalent reformulations of the
Wightman axioms. One is due to Wightman, who wrote down a set of
conditions governing a sequence of tempered distributions
\[
    {\cal W}_n \in {\cal S}(\R^{4n}), \quad
    n = 0,1, \ldots
\]
and proved that, under these conditions, the distributions ${\cal
W}_n$ arise as vacuum expectation values of a unique quantum field
theory satisfying the Wightman axioms, and conversely that the
postulates hold in any Wightman field theory. This is what is
known as the \emph{Wightman reconstruction theorem}, and first
appeared in the seminal paper \cite{Reconstruction}. The part of
this construction relevant to representation theory is known in
functional analysis as the \emph{GNS construction}. A second
reformulation in terms of the Schwinger functions, not directly
used in the present work, was given by Osterwalder and Schrader
\cite{Osterwalder:dx}.

Borchers reformulated Wightman's reconstruction theorem in several
important papers \cite{Borchers:cu,Borchers}, with the result that
a scalar boson quantum field theory is known to be characterized
by a topological \stalg $A$ (with unit element $1_A$) and a
continuous positive form $\omega$ on $A$, satisfying
\be \lab{defstate}
    \omega(aa^*)\geq 0,
    \qquad
    \omega(1_A)=1,
    \qquad
    a\in A.
\ee
A functional satisfying \er{defstate} is called a \emph{state}.

Realistic models are generally described by tensor algebras, and
the action of the state $\omega$ is computed from vacuum
expectation values of products of fields. Although there is reason
to believe the framework of states on tensor algebras could apply
in general to a large class of quantum field theories, previous
formulations of the Wightman reconstruction theorem have focused
on scalar boson quantum field theories.

There are by now many known examples of a low-energy limit or
compactification of string theory which are equivalent to a gauge
theory with compact gauge group. It is also known that exact
quantum string amplitudes can be computed from various flavors of
matrix models. If a mathematically rigorous description using
constructive field theory is possible for these problems coming
from string theory, then a consistent generalization of the
Wightman reconstruction theorem which incorporates the structure
of gauge theory and matrix models seems a useful framework in
which to formulate the result.

The purpose of the present paper is to extend the work of Wightman
and Borchers to include matrix-valued fields of the type required
by gauge theory and matrix models. We first develop the
mathematics, and then make contact with physical applications. The
remainder of this introduction reviews the well-known Wightman
procedure for commuting scalar fields; this serves to fix notation
and set the context for the later sections. Section
\ref{sec:NCtarget} presents the main new idea of the paper, a
generalization of the Borchers construction, and considers some
simple examples. Section \ref{sec:gaugetheory} is concerned with
the application of these ideas to two-dimensional Yang-Mills
theory. In Section \ref{matrixmodels} we recall important recent
work which applies matrix models to high energy physics and then
show that, in the same sense in which scalar quantum field
theories are Wightman states, matrix models are described by
\emph{matrix states}, and thus are special cases of the
construction in Section \ref{sec:NCtarget}. Moreover, matrix
states form a dense subset of the space of all states, and hence
arbitrary field theories are given as limits of matrix models. The
conclusion is that traditional constructive quantum field theory,
gauge theories, matrix models and certain hypothetical
generalizations of these may all be described within a unified
algebraic framework.

\subsection{The Borchers Construction} \label{sec:borchers}

Let
\[
    \s_0 = \C,
    \quad
    \s_n = \s(\R^{kn}),
    \quad \te{and} \quad
    \S = \bigoplus_{n=0}^\infty \s_n.
\]
The latter is a complete nuclear space under the direct sum
topology. There is a natural map $\iota : \otimes^k \s(\R^n) \to
\s(\R^{kn})$ given by $\iota(f_1 \otimes \ldots \otimes f_k) =
\prod_{j=1}^k f_j(x_j)$, where each $x_j \in \R^n$, and the image
of $\iota$ is dense.

Endow $\S$ with the noncommutative multiplication
\bea
    (f \times g)_l &=& \sum_{i+h=l} f_i \times g_h,
    \lab{eq:borch-mult}    \\
    f_i \times g_h(x_1, \ldots, x_{i+h})
    &=&
    f_i(x_1, \ldots, x_{i}) g_h(x_{i+1}, \ldots, x_{i+h}) \nn
\eea
and the involution $f^* = (f_0^*, f_1^*, \ldots)$, where $f_0^* =
\ol{f_0}$ and for $i \geq 1$,
\be  \label{eq:borch-invol}
f_i^*(x_1, \ldots, x_{i}) = \ol{f_i(x_i, x_{i-1}, \ldots, x_{1})}
\ee
The multiplication $\times$ and the unit $\I = (1,0,0,\ldots)$
make $\S$ into a unital $\ast$-algebra with no zero divisors. The
center of $\S$ is $\{ \lambda \I : \lambda \in \C\}$, $\I$ is the
only nonzero idempotent, and the set of invertible elements equals
the center. This implies the triviality of the radical
\[
    \rad(\S)
        =
    \{ g \in \S : \I + f \times g
    \te{ has inverse } \forall \ f\} \, .
\]

An element $g \in \S$ is called \emph{positive} if $\exists f_i$
such that $g = \sum_i f_i^* \times f_i$. This induces a positive
cone $\S^+$ and a semi-ordering. We define the set of
\emph{Hermitian elements}
\[
    \S_h = \{ f \in \S : f^* = f \},
\]
which is a real vector space and we have $\S = \S_h + i \S_h$.
Also, $\S^+$ is a convex cone with $\S^+ \cap (-\S^+) = \{0\}$.
Moreover, we have $\S_h = \S^+ - \S^+$, which follows by
polarization.

\subsection{The Wightman Axioms} \label{sec:wightax}

Let $\s$ denote an appropriate space of test functions, often
taken to be ${\cal S}(\R^d)$.

\begin{axiom} \label{ax:existence}
There exists a Hilbert space $\H$ and a dense domain $D \subset
\H$ such that for every $f \in \s$, an operator $\vp(f)$ exists,
such that $D \subset \dom(\vp(f))$, $\vp(f)D \subset D,$
\[
    (\psi, \vp(f)\chi)
        =
    (\vp(\ol{f})\psi, \chi) \ \te{ for all } \
    \psi, \chi \in D
\]
and $f \to (\psi, \vp(f)\chi)$ is a continuous linear functional
on $\s$.
\end{axiom}

\begin{axiom}
Let $f_a(x) = f(x-a)$. There exists a strongly continuous unitary
representation $\U$ of the translation group $\SymmetryGp$, such
that for all $a \in \SymmetryGp$,  $\U(a)D \subset D$ and
\[
    \U(a) \vp(f) \U^{-1}(a)\psi = \vp(f_a)\psi
\]
for all $f \in \s, \psi \in D$.
\end{axiom}

In standard constructive quantum field theory models, there is a
canonical action of the proper orthochronous Poincar\'e group
$P_+^\up$ on $\s_n$ for all $n$, in other words a representation
$\alpha : P_+^\up \to \Aut(\S)$. We mention the representation
$\alpha$ because even in the generalized models to be introduced
in Section \ref{sec:NCtarget}, invariance under a symmetry group
is expressed in terms of representations similar to $\alpha$. See
also Section \ref{sec:conclusions}.

\begin{axiom}
There exists $\Omega \in D$ such that $\U(a)\Omega = \Omega$ for
all $a \in \SymmetryGp$, and the set of vectors of the form
$\{\Omega,~ \vp(f) \Omega,~ \vp(f_1)\vp(f_2) \Omega,~ \ldots \}$
spans $\H$.
\end{axiom}

Further Wightman axioms will be discussed in Section
\ref{moreaxioms}.

\subsection{States, the GNS Construction, and Axiom \ref{ax:existence}}

Let $\S'$ be the space of continuous linear functionals $T : \S
\to \C$. For $f \in \S$, denote the action of $T$ by $(T,f)$. The
space $\S'$ also has a natural involution; define $T^*$ by
\[
    (T^*,f) := \ol{(T, f^*)}
    \, .
\]
We say a functional is \emph{real} if $T = T^*$, and
\emph{positive} if $(T,p) \geq 0$ for all $p \in \S^+$. The
corresponding spaces are denoted $\S_h'$ and ${\S'}^+$.

The \emph{set of states} is
\[
    E(\S) = \{T \in {\S'}^+ : (T,1)=1 \}.
\]
The \emph{left-kernel} of a state $T$ is defined to be
\be \label{leftkernel}
    L(T) := \{ f \in \S : (T, f^* \times f) = 0\}.
\ee
The left-kernel is so named because it is a left ideal in the
Borchers algebra. The \emph{right-kernel} $R(T)$, defined by the
analogous relation $(T, f \times f^*) = 0$, is a right ideal.
These kernels arise in the quantization procedure discussed later;
choice of the left-kernel amounts to the convention that a
sesquilinear form is conjugate-linear in the first variable.

\begin{theorem} \label{thm:gns}
Each state $T \in E(\S)$ canonically defines a representation
$A_T$ of $\S$ in a Hilbert space $\H_T$ such that the restriction
$A_T |_{\s_1}$ satisfies Axiom 1. Conversely, if $~\{\phi(f)\}~$
are a set of fields satisfying Axiom 1, then every $\Omega \in D$
defines a continuous linear functional $T_\Omega$, by
\[
    (T_\Omega, f_1 \times f_2 \times \dots \times f_n)
    =
    (\Omega, \phi(f_1)\phi(f_2) \ldots \phi(f_n)\Omega),
    \quad
    f_i \in \s_1
\]
If $\| \Omega \| = 1$, then $T_\Omega$ is a state. The field
$A_{T_\Omega}$ is unitarily equivalent to $\{A_\Omega, D_\Omega,
\H_\Omega \}$ where
\[
    D_\Omega
    =
    \te{Linear Span of} ~~ \Omega, ~\phi(f) \Omega,~ \phi(f_1)\phi(f_2)\Omega,~
    etc.
\]
$\H_\Omega$ is the closure of $D_\Omega$, and $A_\Omega(f) =
\phi(f)\big|_{D_\Omega}$.
\end{theorem}

\vskip 0.1 in

\begin{proof}
As a full proof can be found elsewhere
\cite{Borchers:cu,Streater:vi}, we merely recall the central idea
for convenience, as it is used later. $T$ defines a non-degenerate
positive definite sesquilinear form on $\S / L(T)$ by the relation
\[
([f], [g]) = (T, f^* \times g), \ \ [f],[g] \in \S/L(T)
\]
Define $\H_T$ to be the completion of the pre-Hilbert space $\S /
L(T)$, and define a representation of $\S$ by $\phi(f)[g] = [f
\times g]$ for $f \in \s_1$ and $g \in \S$. The rest of the proof
is straightforward.
\end{proof}

\subsection{Translation Invariant States Satisfy Axioms 1-3}

Let $a \in \R^4$. The map $\alpha_a$ defined by
\[
    \alpha_a f_i (x_1, \ldots, x_i)
        =
    f_i(x_1-a, \ldots, x_i-a)
\]
is an element of $\Aut(\S)$. A state $T$ is
\emph{translation-invariant} if $(T, \alpha_a f) = (T,f)$ holds
for all $f \in \S,$ and for all $a \in \R^4$.

\begin{theorem}
Let $T$ be a translation invariant state. Then $A_T(f)$ satisfies
Axioms 1-3. Conversely, if the system $\{A(f), D, \Omega\in D\}$
satisfies Axioms 1-3, then $T_\Omega$ defined by $(T_\Omega, f) =
(\Omega, A(f) \Omega)$ is translation invariant.
\end{theorem}

\subsection{Tensor Products of States}

Given two states $T_1, T_2 \in E(\S)$, let $\{ A_i(f), \H_i, D_i,
\Omega_i \}$ be the associated GNS representations. Then the
triple
\[
    \{ A_1(f) \ox I_2 + I_1 \ox A_2(f),
      \ \H_1 \ox \H_2,
      \ D_1 \times D_2 \}
\]
satisfies Axiom 1, and hence it corresponds to a new state, $T_1
\os T_2$ which is the same as the vector state $T_\Omega$ with
$\Omega = \Omega_1 \times \Omega_2$. Let $P_{n,m}$ denote the set
of all ordered splittings of $n+m$ elements into two subsets, of
respective sizes $n$ and $m$. Let $T_n \in \S_n'$, $S_m \in
\S_m'$, then $T_n \os S_m$ is given by
\[
    (T_n \os S_m)(x_1, \ldots, x_{n+m})
    =
    \sum_{P_{n,m}} T_n(x_{i_1},\ldots, x_{i_n})
    S_m(x_{j_1}, \ldots, x_{j_m})
\]
For $T,S \in \S'$, we define $(T \os S)_n = \sum_{i+k=n} T_i \os
S_k$. This coincides with our previous definition of the
$\os$-product. It is clearly associative and abelian.

\subsection{Real Scalar Fields}

Before discussing more complicated generalizations, we briefly
indicate how the above construction can describe the salient
properties of the quantum theory of a one-component real scalar
field.

In quantum theory of real scalar fields on $\R^m$, the field
algebra is the Borchers algebra $A_\Phi$ where $\Phi =
\mathbb{R}S_4$, the real subspace of the Schwartz space ${\cal
S}(\R^m)$ of complex $C^\infty$ functions $f(x)$ on ${\R}^m$ such
that
\[
    \|\phi\|_{k,\,l}
    ~=~
    \max_{|\al|\leq l} ~
    \sup_{x\in{\R}^m}
    (1+|x|)^k \frac{\d^{|\al|}}{(\d x^1)^{\al_1}
    \ldots(\d x^m)^{\al_m}} f(x) < \infty
\]
for any collection $(\al_1,\ldots,\al_m)$ and all $l,k\in \N$. The
space ${\R}S_m$ is endowed with the set of seminorms
$\|\phi\|_{k,\,l}$ and the associated topology. It is reflexive.

Since $\otimes^k {\cal S}(\R^n)$ is dense in ${\cal S}(\R^{kn})$,
every continuous form on the subspace has a unique continuous
extension. Every bilinear functional $M(\phi_1, \phi_2)$ which is
separately continuous in $\phi_1\in {\cal S}(\R^n)$ and $\phi_2\in
{\cal S}(\R^m)$ may be expressed uniquely in the form
\[
    M(\phi_1,\phi_2)
    =
    \int F(x,y)\phi_1(x)\phi_2(y)\, d^n x\, d^m y, \qquad
    F\in {\cal S}'(\R^{n+m}).
\]

As a consequence, every state $\omega$ on the Borchers algebra
$A_{\R S_4}$ is represented by a family of distributions $W_n \in
{\cal S}'(\R^{4n})$. For any $\omega$, there exists a sequence
$\{W_n \}$ such that
\be \label{om-int}
    \omega(\phi_1 \times \cdots \times \phi_n)
    =
    \int
    W_n(x_1,\ldots,x_n)
    \phi_1(x_1) \cdots \phi_n(x_n)\,
    d^4x_1\cdots d^4x_n.
\ee

A rigorous proof is known that a scalar quantum field theory, when
it exists, is completely determined by its Wightman functions. By
relation \er{om-int}, a state $\omega$ on the Borchers algebra
contains the same information as a complete specification of the
$n$-point functions for all values of $n$. If $\omega$ obeys the
Wightman axioms, then we have a quantum field theory and $W_n$ are
related to vacuum expectation values of products of fields. They
calculate observable quantities such as cross-sections and decay
rates.

\subsection{Spectral Condition, Locality, and Uniqueness of the
Vacuum}\label{moreaxioms}

The remaining two essential properties of a quantum field theory
(spectral condition and locality) are equivalent to $\ker \omega$
containing certain ideals.

Let $\s_1(CV^+)$ denote the set of functions in $\s_1$ that vanish
on the forward light cone $V^+$, and let ${\cal F}$ denote the
Fourier transform. The spectral condition is the statement that
$\ker \omega \supset I_1$, where
\[
    I_1 =
    \left\{ \int d^4a F(a) \alpha_a f
    ~:~
    f \in \S,~ f_0 = 0,~ F(a) \in {\cal F} [\s_1(CV^+)]
    \right\}
\]

Spacetime locality is the statement that $\ker \omega \supset
I_2$, where $I_2$ is the smallest closed two-sided ideal in $\S$
containing all elements of the form $f \times g - g \times f$
where $f$ and $g$ have spacelike-separated supports.

Uniqueness of the vacuum also has a simple interpretation in terms
of Wightman functionals. A field theory is said to be
\emph{reducible} if the algebra of field operators acts reducibly
on the Hilbert space. A Wightman state $\omega$ is said to be
\emph{decomposable} if there exists a positive number $\la < 1$
such that
\be \lab{decompos}
    \omega = \la \omega^{(1)} + (1-\la) \omega^{(2)}
\ee
with Wightman states $\omega^{(1)}$ and $\omega^{(2)}$ different
from $\omega$. Indecomposability of the Wightman functional is
equivalent to uniqueness of the vacuum in an irreducible field
theory.

In Section \ref{sec:NCtarget} we will generalize the Wightman
state $\w$, and it is of interest to know whether \er{decompos}
also leads to uniqueness of the vacuum in the general case.

\section{Noncommutative Target Space Perspective}
\label{sec:NCtarget}

The field algebra with multiplication and involution given by
\eqref{eq:borch-mult}-\eqref{eq:borch-invol} admits a natural
generalization to the noncommutative setting. This generalization
has many applications in physics, all of which come from
interpreting elements of the Borchers algebra as gauge fields on a
$d$-dimensional spacetime. For $d \leq 1$, this gives rise to
matrix models and matrix quantum mechanics. For $d \geq 2$, it is
Yang-Mills theory. Of course, the gauge symmetry is not essential
for the construction to work; it applies equally well to
matrix-valued scalar field theory of the type considered by
Kazakov \cite{Kazakov:2000ar}. This framework is also suggestive
of quantum field theory in which the target manifold is a
noncommutative space in the sense of Connes.

\subsection{Test Functions Valued in a Noncommutative Space}

First we wish to argue that a correct description of gauge quantum
field theory is possible in terms of test functions valued in a
noncommutative space. Quantum fields are operator-valued
distributions. Consider a pure gauge theory with gauge group $G$
and Lie algebra $\g = \Lie(G)$. In a classical pure gauge theory,
the fundamental fields are $\g$-valued one-forms, each determining
a connection on a principal $G$-bundle. In a quantum version of
the same gauge theory, these classical fields would be promoted to
operator-valued distributions with the same algebraic structure.

For concreteness, let ${\cal S}(\R^4)$ denote the Schwartz space
of rapidly decreasing functions on $\R^4$. An operator-valued
distribution is a continuous map
\[
  {\cal S}(\R^4) \To \Op(\H)
\]
where we consider ${\cal S}(\R^4)$ to be endowed with the Schwartz
topology, $\H$ is a Hilbert space, and $\Op(\H)$ denotes a
suitable space of unbounded operators on $\H$. In the example
of a free real scalar boson, $\H$ is the usual bosonic Fock
space, and $\Op(\H)$ would be a class of operators large enough
to include all operators of the form $\phi(f)$, where $\phi$ is
a quantum field and $f$ is any
test function. The operators $\Op(\H)$, in this example, have
a common core including all smooth, compactly supported Fock
states with finite particle number.

Let $\rho : \g \to \gl(V)$ be a representation of $\g$ on the
representation space $V$. Let $\H^V$ denote the space of all
continuous linear functionals
\be \label{distribution-lie}
  \phi : {\cal S}(\R^4) \To \Op(\H) \otimes V \, .
\ee
Elements of $\H^V$ are operator-valued distributions that
transform in the representation $V$. The representation $\rho$ on
$V$ naturally defines a representation $\bar{\rho}$ of $\g$ on
$\H^V$ by expanding
\[
    \phi(f) = \sum_{i=1}^n A_i^f \otimes v_i^f
\]
for $A_i^f \in \Op(\H), v_i^f \in V$, and defining
\[
    (\bar{\rho}(g) \phi)(f) = \sum_i A_i^f \otimes
    (\rho(g)v_i^f), \quad
    g \in \g \, .
\]
One very useful choice for $V$ is the adjoint representation,
because field strength variables $F_{\m\n}(x)$ in Yang-Mills
theory transform in the adjoint. Other representations typically
arise as direct summands of tensor powers of the adjoint and its
complex conjugate representation. In all of these examples, it is
useful to view the representation space $V$ as living in some
matrix algebra.

There is a natural transformation of categories by which the space
$\H^{V}$ defined above is naturally isomorphic to the space
$\H_{V^*}$ of continuous maps
\be \label{dual}
   {\cal S}(\R^4)\otimes V^* \To \Op(\H)
\ee

Let us exhibit the isomorphism between \er{distribution-lie} and
\er{dual} explicitly. For concreteness, we will fix our attention
on the special case of $V = \g$, the adjoint, but we stress that
no part of the discussion depends on this in an essential way.

In terms of $\phi$ we may define a new map
\be \lab{tilde}
  \til{\phi} : {\cal S}(\R^4) \times \g^* \To \Op(\H)
\ee
by the formula
\be \label{tilphi}
  \til{\phi}(f, y) \equiv
  \sum_{i=1}^n y(v_i^f) A_i^f \,,
  \quad
  y \in \g^*
  \ .
\ee
Equivalently, if $T^a$ is a basis for $\g$, and $\phi(f) =
\phi(f)_a T^a$, then for $y \in \g^*$, $\til\phi(f, y^a) =
\phi(f)_a y^a$.

The map $\til{\phi}$ is multilinear, and therefore factors through
to a map on the tensor product ${\cal S}(\R^4) \otimes \g^*$. We
have proved that $\phi \to \til{\phi}$ gives an explicit
isomorphism $\H^V \cong \H_{V^*}$ between the two spaces
\er{distribution-lie} and \er{dual}.

Quantum field theory with test functions taking values in $\g^*$
is most naturally described by a noncommutative version of the
Borchers construction, and the latter mathematical structure will
occupy us for the rest of this section and, in some form or other,
for the rest of the paper. We summarize the results of the
previous paragraphs in a lemma.

\begin{lemma}The following structures are equivalent:

\begin{enumerate}
\item An operator-valued distribution which transforms in the
adjoint representation of a Lie algebra $\g$ (i.e. a quantized
Yang-Mills field)

\item An operator-valued distribution which acts on $\g^*$-valued
test functions.
\end{enumerate}
\end{lemma}

Let us see how this isomorphism works in practice. Suppose that
$F_{\m\n}(x)$ is an operator-valued distribution which transforms
as a Lie algebra-valued two-form. An example of such an object is
a quantized Yang-Mills field strength. The above construction
tells us that from $F_{\m\n}(x)$, we can construct a single
operator-valued distribution $\til F_{\m\n}$ which acts on test
functions $f(x)$ valued in the dual of the Lie algebra. The
duality between $\g$ and $\g^*$ is given explicitly by the Killing
form $K(a,b) = \tr(ab)$, where the trace is taken in the adjoint
representation. Therefore, the correct definition is
\be \label{Ftilde}
    \til F_{\m\n}(f)
    = \tr(f \cdot F_{\m\n})
    = \int K\big( f(x), \, F_{\m\n}(x) \big)\, dx
\ee
where $f \cdot F_{\m\n}$ is defined by
\[
    [f \cdot F_{\m\n}]_{ij}
    \ \equiv \
    \sum_k \int f(x)_{ik} (F_{\m\n}(x))_{kj} \, dx
\]
The notation of \er{Ftilde} is the same as \er{tilde}. This shows
explicitly how operator-valued distributions act on $\g^*$-valued
test functions.

\begin{remark} \label{remark:vectorbundle}
This duality transformation transfers the dependence on the Lie
algebra to the test functions; however, if the original field also
transforms as a section of an additional vector bundle $E$, as is
the case for the 2-form $F_{\m\n}(x)$ which is a section of $E =
\wedge^2(\R^d)$, the quantized field operator \er{Ftilde}
transforms in the tensor product $E \otimes \O_{\H}$, where
$\O_\H$ is a trivial bundle with fiber $\Op(\H)$. Additional
complications to the theory presented above arise in the case of a
nontrivial fibre bundle, in which the connection can be only
\emph{locally} described as a $\g$-valued one-form. The gauge
fields in the present paper are all assumed to be sections of
globally trivial principal bundles.
\end{remark}

\begin{remark}
One could imagine exotic quantum field theories in which the
fields take values in the algebra of functions over a
noncommutative geometry in the sense of Connes. The present
constructions define such a theory mathematically, but we do not
know a direct physical interpretation.
\end{remark}

\subsection{Generalizing the Borchers Construction} \label{core}

For any space $\Sigma$ let
\be \label{GBtestfn}
    \A = \A(\Sigma, \B)
\ee
denote a vector space of ``test functions'' from $\Sigma$ to a
possibly noncommutative star-algebra $\B$ with product $~\cdot~$.
If $\B$ is a normed algebra and $\Sigma$ is equipped with an
appropriate metric, then we may consider $\A(\Sigma, \B)$ to be
the Schwartz space of rapidly-decreasing functions. For example,
one could consider a spacetime which has nontrivial topology
within some compact region $K$, and outside that region it is
covered by a single chart and approximately isometric to $\R^n$
minus a compact set. On such a spacetime, one can demand that the
$\B$-norm of the function and of all its derivatives, expressed in
any chart which covers the complement of $K$, fall off faster than
any power of the \emph{modulus function}, which measures the
distance of a point from $K$ in the ambient metric.

In the application of these ideas to two-dimensional gauge theory,
$\Sigma$ will denote a compact Riemann surface. We assume for
convenience that the base field of $\B$ is $F = \R$ or $\C$. $\A$
is then naturally a left and right module over $\B$, and of course
also over $F$. Let
\be \lab{def-Abar}
    T\A = \bigoplus_{k=0}^\infty \T^k\A,
    \quad
    \te{ where }
    \quad
    \T^k\A = \A^{\otimes k}
\ee
We will abbreviate $\A \otimes \dots \otimes \A$ ($n$ factors) by
$\A^{\otimes n}$. Let $x$ and $y$ denote elements of $\Sigma$. We
define a map $\iota$ which identifies $f \otimes g$ with the
$\B$-valued function of two variables given by $f(x) \cdot g(y)$.
This identification, and the natural extension of this map to
higher tensor powers $\A^{\otimes n}$, determine an algebra
homomorphism
\be \label{inj-1}
\xymatrix{
    {\A^{\otimes n} \ }
    \ar[r]^{\iota\ \ \ } &
    {\ \A(\Sigma^n, \B)}
    }
\ee
where $\Sigma^n$ denotes the $n$-fold Cartesian product $\Sigma
\times \Sigma \times \ldots  \times \Sigma$. In other words,
\[
    \iota(f_1 \otimes \dots \otimes f_n)(x_1, \ldots, x_n)
    ~=~ f_1(x_1) \cdot \ldots \cdot f_n(x_n)
    ~.
\]
Both the kernel and the image of this homomorphism are important.

The map $\iota$ can have a nonzero kernel if $f(x)$ commutes with
$g(y)$ for all $x,y \in \Sigma$, in which case $\iota(f \otimes g)
= \iota(g \otimes f)$. For our purposes, we would like to assume
that \eqref{inj-1} is injective; to attain this injectivity it is
necessary to quotient by the kernel of $\iota$, which is
equivalent to working with the universal enveloping algebra as we
now discuss.

Since $\B$ is defined to be an associative algebra, it is
naturally also a Lie algebra. Therefore the algebra $\A$ of
functions $\Sigma \to \B$, is also a Lie algebra. Let $\uA$ be the
universal enveloping algebra ${\cal U}(\A) = T\A / I$ where $I$ is
the two-sided ideal generated by elements of the form
\[
    f \otimes g - g \otimes f - [f,g],
    ~~~~~
    f, g \in \A
\]
Since $I = \ker(\iota)$, it follows that $\iota$ is injective on
$\uA$.

This subtlety is only present in the noncommutative version, and
leads to another subtlety in the definition of the grading. For
the symmetric tensor algebra, which corresponds to the abelian
case $[\cdot, \cdot] = 0$, the grading $T\A = \bigoplus_k \T^k$
projects to a grading on $\U(\A)$, but in general projection of
$T\A$ to $\U(\A)$ does \emph{not} produce a grading. However,
$T\A$ has the associated filtration
\[
    \T^{(k)} = \bigoplus_{j=0}^k \T^j \A
\]
from which we recover $\T^k$ by
\[
    \T^k \simeq \T^{(k)} / \T^{(k-1)}
\]
Let $\U^{(k)}$ denote the image of $\T^{(k)}$ under the
projection. This defines a natural filtration,
\[
    \U^{(k)} \U^{(\ell)} \subseteq \U^{(k+\ell)} ,
\]
and the spaces $\A_k = \U^{(k)} / \U^{(k-1)}$ provide the desired
grading of $\uA = \U(\A)$. We will often write $f = \{ f_0, f_1,
f_2, \ldots \}$ to denote the decomposition of an element $f \in
\uA$ with respect to this natural grading.

If $\B$ is a real, abelian algebra, then the above construction
reduces to the classic construction of Wightman. In particular, we
may consider $\B = \R$ which corresponds to a single real scalar
field. In this case, $[f,g] = 0$ always, hence $I$ is the ideal
generated by elements of the form $f \otimes g - g \otimes f$.
This identifies $\uA$ with the symmetric tensor algebra over $\A$,
which is precisely the algebra used in Wightman's original
construction \cite{Streater:vi}. Once again, $\iota$ is injective
on the space of interest; this injectivity is an important
component of the Wightman construction.

When the interpretation is clear from context, as in
\eqref{cross-1}, we will not explicitly write the map $\iota$.
With this convention, our notation becomes compatible with the
notation of \cite{Streater:vi}, in which (for example) one would
write
\[
    (h \otimes f_k)(x_1, \ldots, x_{k+1})
    =
    h(x_1) f_k(x_2, x_3, \ldots, x_{k+1})
\]

Since $\uA$ is generated as a vector space by homogeneous
elements, we can define a cross product on $\uA$ by the equation
\be \label{cross-1}
    (f_n \times g_m)(x_1, \ldots, x_{n+m})
    =
    f_n(x_1, \ldots, x_n)
    \cdot
    g_m(x_{n+1}, \ldots, x_{n+m})
\ee
For $f_n \in \A(\Sigma^n, \B)$ and $g_m \in \A(\Sigma^m, \B)$,
this determines an element
\[
    f_n \times g_m \in \A(\Sigma^{n+m}, \B),
\]
and this element is the same as the function $\iota(f_n \otimes
g_m)$. The cross product extends to all of $\uA$ in a manner
similar to eq.~\eqref{eq:borch-mult},
\be \label{cross-2}
    (f \times g)_l = \sum_{i+h=l} f_i \times g_h \, .
\ee

The involution on $\uA$ is defined as follows. For $f \in \A$, we
define $f^*(x) := f(x)^{*_{\B}}$ where ${*_{\B}}$ denotes the
involution in $\B$. We define the involution to satisfy the axioms
of a $\ast$-algebra, so that $(f + \lambda h)^* = f^* +
\bar\lambda h^*$ and $(f \times g)^* = g^* \times f^*$. Since $\A$
generates $\uA$ as an algebra, this determines the star operation
on all of $\uA$.

In the above, $\B$ was defined to be an associative algebra, and
we also used the natural Lie bracket coming from commutators in
the associative algebra's product. It is worth noting that a
completely analogous construction may be carried out even in the
case where $\B$ is an abstract Lie algebra on which no associative
algebra structure is defined. To do this, the definition of $\uA$
as the universal enveloping algebra is the same, and the
definition of the cross product \eqref{cross-1} could be modified,
to become
\be
    (f_n \times g_m)(x_1, \ldots, x_{n+m})
    =
    \big[ f_n(x_1, \ldots, x_n)
    , \
    g_m(x_{n+1}, \ldots, x_{n+m})
    \big]
\ee
However, this is unlikely to be a useful construction because the
actual Hilbert space inner product will be identically zero,
assuming the state $t \in E(\B)$ used there is a cyclic state,
meaning that
\[
    t(ab \ldots c) = t(cab\ldots) \ .
\]
Of course, the trace on any matrix algebra or Hilbert space is a
cyclic state.

\subsection{States, Sesquilinear Forms, and Field Operators}

The cross product provides a mapping from states to the bilinear
forms that arise in quantum physics. Explicitly, the advantage of
the cross product \eqref{cross-1}-\eqref{cross-2} is that any
state $\omega$ on $\uA$ determines a sesquilinear form $\< \ , \
\>_\omega$ by the relation
\be \label{bil-form}
    \< f, g \>_\omega = \omega(f^* \times g),
\ee
and the sesquilinear forms \eqref{bil-form} are of the type that
arise in the construction of the Fock-Hilbert space for a quantum
field theory. The left-kernel of the state $\omega$ is precisely
the set of $f$ such that $\<f,f\>_\omega = 0$. Since a state by
definition satisfies the positivity axiom, the associated
sesquilinear form
    $\< \ , \ \>_\omega$
is a positive semi-definite inner product.  It is a positive
definite inner product on the quotient by the left-kernel $L(\w)$
of the state $\omega$, and the completion of $\uA / L(\w)$ forms
the physical Hilbert space $\Hphys$.

The state $\omega$ defines a non-degenerate positive definite
sesquilinear form on $\uA / L(\omega)$ by the relation
\be \lab{eq:sesqui}
    ([f],[g]) = \omega(f^* \times g)
\ee
Define field operators $\Phi(f), f \in \A$ acting on $\uA$ by the
formula
\be \lab{fieldoperator}
    \Phi(f)(a_0, a_1, a_2, \ldots)
    =
    (0, f a_0, f \otimes a_1, f \otimes a_2, \ldots )
\ee
Define $\Hphys$ to be the completion of the pre-Hilbert space $\uA
/ L(\omega)$, and define a representation of $\A$ on $\Hphys$ by
\be \lab{vp}
    \vp(f)[g] = [f \times g]
\ee
for $f \in \A,\ g \in \uA$. A short proof shows that $\Phi$, as
defined by \er{fieldoperator}, is well-defined on equivalence
classes and is the same as $\vp$ upon passing to the quotient.

\subsection{Defining a State from Wightman Functions}

In the generalized Borchers construction, a sequence $\W = \{ W_n
\}$ of tempered distributions does not directly define a
(complex-valued) sesquilinear form. This represents a departure
from the usual quantum theory of scalar fields. The sequence $\W$
does naturally define an associated linear map $\Omega_\W : \uA
\to \B$ by
\[
    \Omega_\W(f = \{f_0, f_1, \ldots \})
    =
    \sum_{n=0}^\infty \int f_n(x_1, \ldots, x_n) \cdot W_n(x_1, \ldots, x_n)
    \ {\textstyle \prod\limits_{i=1}^n dx_i}
    \ \in \ \B
\]
The sum in this expression is always well-defined, since $f$, by
assumption, has finitely many nonzero components.

In the intended application, a scalar-valued functional $\omega$
is recovered by composing $\Om_\W$ with a natural scalar-valued
state $\tr : \B \to \C$, given by the trace. The functional
$\omega$ then defines a sesqui-linear form via \eqref{eq:sesqui}.
Let us use the term \emph{pulled-back states} for those which
arise from composing the linear map $\Omega_\W$ with a state on
$\B$. The states which arise in reconstruction theorems for gauge
theories necessarily take this form.

Every state $t$ on $\B$ determines a sesqui-linear form $\< \ , \
\>_{HS}$ on $\B$, defined by
\be \lab{HS}
    \< a , b \>_{HS} = t(a^* \cdot b)
\ee
which we call the \emph{Hilbert-Schmidt form}. In case $\B$ is a
$C^*$-algebra represented by bounded operators on a Hilbert space
$\H$, the trace $\Tr_{\H}(\ )$ is the natural state to use,
\er{HS} is the usual Hilbert-Schmidt inner product, and
consequently \er{HS} determines a Hilbert space structure on $\B$.
The Hilbert-Schmidt inner product arises often when we construct
examples of noncommutative Wightman states.

\subsection{The Action of a Symmetry Group}

In this picture, the action of a symmetry group $\SymmetryGp$
(such as, for example, Lorentz symmetry which acts on test
functions by transforming the spacetime point on which the test
function is evaluated) would be most easily defined at the level
of the field algebra by an $\omega$-invariant representation
$\alpha : \SymmetryGp \to \Aut(\uA)$. The condition of
$\omega$-invariance is already a stronger condition than classical
symmetry, since $\omega$ contains all quantum correlation
functions of the theory. However, in order to obtain a full
symmetry of the quantum theory, we must require additionally that
for all $g \in \SymmetryGp$, we have $\alpha_g (L(\omega))
\subseteq L(\omega)$ so that the action of the symmetry descends
to a representation on the physical Hilbert space $\Hphys$, which
we will call $\hat\alpha$. It follows from the fact that $\alpha_g
\in \Aut(\uA)$ and by $\omega$-invariance that for any $g$,
$\hat\alpha_g$ is unitary on $\Hphys$.

A deep and beautiful question in quantum field theory involves
whether or not a classical symmetry is preserved quantum
mechanically; if not the symmetry is said to be
\emph{spontaneously broken}. If the quantum mechanical ground
state $\Om$ is uniquely given by the equivalence class in $\uA /
L(\omega)$ of the unit element $\I \in \uA$, if $\alpha_g$
preserves $L(\omega)$ for all $g$, and if $\alpha_g \I$ is
proportional to $\I$, then we infer that $\hat\alpha_g \Om = \Om$
and the symmetry is unbroken.

\subsection{All Continuous Linear Functionals on the Field Algebra}

In the introduction, we mentioned the well-known result that every
bilinear functional $M(\phi_1, \phi_2)$ which is separately
continuous in $\phi_1\in {\cal S}(\R^n)$ and $\phi_2\in {\cal
S}(\R^m)$ may be expressed uniquely in the form
\[
    M(\phi_1,\phi_2)
    =
    F(\phi_{12}), \qquad
    F\in {\cal S}'(\R^{n+m}),
\]
where
\be \label{phi12}
    \phi_{12}(x_1, \ldots, x_{n+m}) \equiv
    \phi_1(x_1, \ldots, x_n) \phi_2(x_{n+1}, \ldots, x_{n+m}).
\ee
This implies, then, that every state on the Borchers algebra is
determined by a collection of distributions which, assuming the
relevant axioms hold, may be identified with the $n$-point
functions of a quantum field theory.

The purpose of the present section is to prove the analogous
result for the general Borchers algebra introduced in Section
\ref{core}.

\begin{theorem}
Suppose that $\vp_1$ and $\vp_2$ are Schwartz functions from
$\R^n, \R^m$ respectively, into the space of $k \times k$ matrices
over $\C$. Let $\M(\vp_1, \vp_2)$ be a bilinear functional,
separately continuous in both variables. We claim there exists a
distribution $\F \in {\cal S}'(\R^{n+m}, \Mat_{k\times k}(\C))$
such that
\[
    \M(\vp_1, \vp_2) = \F(\vp_{12})
\]
where
\be \label{p12}
    \vp_{12}(x_1, \ldots, x_{n+m})^{ab} \equiv
    \sum_c \vp_1(x_1, \ldots, x_n)^{ac} \vp_2(x_{n+1}, \ldots, x_{n+m})^{cb}.
\ee
\end{theorem}

\begin{proof} $\M$ naturally determines a linear functional $\M|$ on the
subset of ${\cal S}(\R^{n+m},\Mat_{k\times k}(\C))$ consisting of
those matrices of Schwartz functions which are factorizable as in
\eqref{p12}. It is easily seen that the restricted functional
$\M|$ is continuous in the relative topology induced from ${\cal
S}(\R^{n+m},\Mat_{k\times k}(\C))$, and therefore it may be
extended to a continuous map ${\cal F}$ defined on all of ${\cal
S}(\R^{n+m},\Mat_{k\times k}(\C))$.
\end{proof}

We conclude that any state on the generalized Borchers algebra
determines a set of generalized $n$-point functions.

\subsection{Two Simple Examples}
\label{example-zero}

Eq.~\er{bil-form} determines a bilinear form on $\uA$, giving $\uA
/ L(\omega)$ the structure of a pre-Hilbert space. In this
section, we compute this bilinear form for two cases in which the
Wightman functions have a very simple structure.

First suppose the Wightman distributions are given by a sequence
of non-negative real constants, $\alpha_n \in \R_{\geq 0}$, times
the identity matrix. We compute
\begin{eqnarray*}
    \Omega_\W(f^* \times g) &=&
    \sum_{k=0}^\infty \sum_{n+m=k}
    \int \alpha_{n+m} f_n(x_n, \ldots, x_1)^* g_m(y_1, \ldots, y_m)
    \prod_{i,j} dx_i dy_j \\
    &=& \sum_{k=0}^\infty \alpha_k \sum_{n+m=k}
    \Big( \int f_n^* \Big) \Big( \int g_m \Big)
\end{eqnarray*}
The associated sesquilinear form is then given by
\be \label{sesq-form-constW}
    (f, g) =
    \sum_{k=0}^\infty \alpha_k \sum_{n+m=k} \< F_n, G_m \>_{HS}
\ee
where capital letters denote integration, i.e. $F_n = \int
f_n(x_1, \ldots, x_n) d^n x$, etc.

If $\B$ is a $C^*$-algebra, then it can be realized as an algebra
of bounded operators on some Hilbert space $\H$. In this case,
$\B$ is itself a Hilbert space with the standard Hilbert-Schmidt
inner product defined by
\[
    \< A, B \>_{HS} = \Tr(A^* B), \quad
    A, B \in \B .
\]
This extends to tensor products of $\B$ in the usual way,
\be \label{HSTP}
    \< A \otimes B, C \otimes D\>_{HS}
    =
    \Tr(A^*C) \Tr(B^*D).
\ee
This induces a Hilbert-Schmidt type inner product on the algebra
$\uA$, and this is an example of the inner product arising from a
state $\omega$ on the generalized Borchers algebra. Since our
fields are $\B$-valued functions, our inner product will be the
integrated version of the Hilbert-Schmidt inner product for
tensors \eqref{HSTP}. The state which generates this inner product
is defined by considering all possible products of delta
functions.

In certain ``ultralocal'' field theories, the correlation
functions can simply be delta functions. Explicitly, consider the
Wightman functions
\begin{eqnarray*}
    W_2 &=& \delta(x_1 - x_2) \\
    W_4 &=& \delta(x_1 - x_4) \delta(x_2 - x_3) \\
    &\vdots& \\
    W_{2n} &=& \prod_{i=1}^n \delta(x_{i+n} - x_{p(i)})
\end{eqnarray*}
where $p$ is the permutation of $\{1,\ldots,n\}$ that completely
reverses the order, with an implied identity matrix after each
delta function. An example using $W_4$ is
\begin{eqnarray*}
    \< f,g \>
    &=&
    \int \Tr(f(x_2, x_1)^* g(x_3,x_4))
    \delta(x_1 - x_4) \delta(x_2 - x_3) \Pi_i dx_i \\
    &=& \int_{\Sigma^2} \< f, g\>_{HS}
\end{eqnarray*}
At first glance, the integrand of this expression looks different
from \eqref{HSTP}. In fact they are the same; to see this, let $f
= f_a \ox f_b$ and $g = g_a \ox g_b$. Then we have
\begin{eqnarray*}
    \int_{\Sigma^2} \< f, g\>_{HS}
    &=& \int_{\Sigma^2}
    \Tr \Big[ (f_a \ox f_b)^* \cdot (g_a \ox g_b)\Big] \\
    &=& \int_{\Sigma^2}
    \Tr \Big[ (f_a^* \cdot g_a) \otimes (f_b^* \cdot g_b)\Big] \\
    &=& \int_{\Sigma^2}
    \Tr (f_a^* \cdot g_a) \Tr(f_b^* \cdot g_b)
\end{eqnarray*}
as desired.

\subsection{Axioms for Nonpositive Theories} \label{sec:hssc}

A number of quantum field theory models are known which do not
satisfy positivity. The general properties of these models are
summarized in the modified Wightman axioms of indefinite metric
QFT \cite{Morchio:1979gt}. In particular, there are two
replacements of the positivity axiom which immediately generalize
to the noncommutative algebraic framework outlined in the present
work.

Albeverio \emph{et al.} \cite{Albeverio:tv} investigated Euclidean
random fields as generalized white noise and remarked that the
Wightman functionals belonging to those fields do not generally
satisfy positivity. Those nonpositive Wightman functionals satisfy
the following weaker condition, known as the Hilbert space
structure condition \cite{Albeverio:sc}.

\vskip 0.1in

\noindent {\bf Axiom (Hilbert space structure condition).} $\ $
{\it There exist seminorms $p_n$ on $\s_n$ such that
\be \lab{hssc}
    |W_{n+m}(f_n^* \otimes g_m)| \leq p_n(f_n) p_m(g_m)
    \quad\te{ for all }\quad
    f_n \in \s_n,\ g_m \in \s_m.
\ee
}

This axiom needs no modification in order to apply to the general
Borchers construction of Section \ref{core}; the $p_n$ are simply
reinterpreted as seminorms on the subspaces $\A_n$ arising in the
grading of the universal enveloping algebra.

A related condition known as the Krein structure condition
\cite{Jakobczyk:zx} is satisfied by the physically important
Gupta-Bleuler formalism for free QED, and has many attractive
features from a mathematical standpoint.

\vskip 0.1in

\noindent {\bf Axiom (Krein positivity).} $\ $ {\it There exists a
dense unital subalgebra $\A_0$ of the Borchers algebra, and a
mapping $\alpha : \A_0 \to \A_0$, such that for all $f, g \in
\A_0$,
\begin{enumerate}
\item $\omega( \alpha^2(f)^* \times g) = \omega(f^* \times g)$;
\item $\omega(\alpha(f)^* \times f) \geq 0$;
\item $\omega(\alpha(f)^* \times g) = \omega(f^* \times \alpha(g))$; and
\item $p_\alpha(f) \equiv \omega(\alpha(f)^* \times f)^{1/2}$ is
continuous in the topology of the Borchers algebra.
\end{enumerate}
}

\vskip 0.1in

In the original paper \cite{Jakobczyk:zx}, it is shown that the
Krein positivity condition is stronger than the Hilbert space
structure condition, is satisfied by free QED, and guarantees the
existence of a majorizing Krein-type Hilbert space structure
associated to the Wightman functions.

It is easily seen that the Krein positivity condition may be
applied to the generalized Borchers algebra of Section \ref{core},
and a state $\omega$ on that algebra, simply by interpreting the
terminology within the new context.

\section{Application to Gauge Theory}
\label{sec:gaugetheory}

We would like to use the structure developed above to express
quantities of interest in gauge theory. The difficulty with this
outlook is that there are different possible choices for complete
sets of observables. It is known that Wilson loop functionals are
a complete set of observables for Yang-Mills theory in any
dimension, but as functionals on the loop space, they cannot be
directly used to generate a state on the generalized Borchers
algebra. Fortunately, in some cases, a complete set of
gauge-invariant correlation functions is available, and they
possess a mathematical structure which is convenient for our
viewpoint in this paper.

\subsection{Complete Sets of Observables}

In any number of dimensions, the Yang-Mills field strength is a
Lie algebra valued 2-form; a special feature of two dimensions is
that in this case the field strength is mapped to a $\g$-valued
scalar field by the Hodge star. This field is denoted $\xi(x)$,
and given explicitly by
\[
    F_{\mu\nu}(x) = \xi(x) \sqrt{g(x)} \eps_{\mu\nu}
\]
The Yang-Mills action in two dimensions is
\be \label{ymactioni}
    S = \f{1}{8\pi^{2}\eps}
    \int_\Sigma \Tr F \wedge \ast F \, ,
\ee
with the trace taken in the fundamental representation for $\g$.
In our convention, the gauge field $A$ is anti-Hermitian. In
terms of $\xi$ the pure Yang-Mills action takes the form
\[
    \int_\Sigma d\mu \Tr(\xi^2)
\]
with the appropriate coupling constant inserted. Here $d\mu =
\sqrt{g(x)}\, d^2 x$ is the Riemannian volume measure on $\Sigma$.

\emph{Field strength correlators} are linear combinations of
objects of the form
\be \label{corr}
    \< \xi^a(x_1) \xi^b(x_2) \xi^c(x_3) \ldots \xi^d(x_n) \>
\ee
Here $\xi(x)$ is a Lie algebra valued scalar field, and $\xi^a,
\xi^b, \ldots$ come from expanding the field with respect to some
fixed basis of the Lie algebra. For example, one could take the
Gell-mann matrices ${\bf t}_a$ as a basis of $SU(3)$ and write
$\xi(x) = \xi^a(x) {\bf t}_a$. Thus, \er{corr} is not a
gauge-invariant correlator. It becomes gauge invariant only after
inserting ${\bf t}_a, {\bf t}_b, {\bf t}_c, \ldots$, summing over
repeated indices, and taking the trace. The rest of this
subsection will be devoted to describing a second type of
correlator, which are sometimes called \emph{$\phi$-field
correlators}.

In two dimensions there are no propagating degrees of freedom
(i.e. no gluons) so the only degrees of freedom come from
spacetimes of nontrivial topology or Wilson loops. Since there are
so few degrees of freedom, there is a very large group of local
symmetries. $YM_2$ is invariant under the group ${\rm
SDiff}(\ST)$ of area preserving diffeomorphisms, which is a
larger symmetry group than local gauge invariance.

The following equivalent action is called the ``first-order
formalism'' because Gaussian integration over $\phi$ gives back
the original action \eqref{ymactioni}.
\[
    Z_{\Sigma}(\eps)
    =
    \int DA \exp{ \left(\f{1}{8 \pi^2\eps}
    \int_{\Sigma} \Tr F \wedge \ast F  \right)}
    =
    \int DA D\phi \ e^{-S(A,\phi)}
\]
where
\be \label{pi}
    S(A, \phi) =
        -\f{i}{4\pi^2}
        \int_\Sigma \Tr (\phi F)
        -
        \f{\eps}{8 \pi^2}
        \int_\Sigma d\mu \Tr \phi^2
        \, .
\ee
Here $\phi$ is a Lie-algebra valued $0$-form; \eqref{pi} shows
that the gauge coupling $e^2$ and total area $a = \int_\Sigma
d\mu$ always enter together, and gives rise to the natural
generalization
\be \lab{invtpoly}
    I =  \int_\Sigma \biggl[ i \tr (\phi F)  + \V(\phi) d\mu\biggr]
\ee
where $\V$ is any invariant function on the Lie algebra $\g$.
Thus, ordinary $YM_2$ is one example of a general class of
theories parameterized by invariant functions on $\g$. It is
natural to restrict to the ring of invariant polynomials on $\g$.
For $G=SU(N)$, this ring is generated by $\tr \phi^k$, so we may
describe the general theory by coordinates $t_{\vec k}$, in terms
of which
\[
    \V= \sum t_{\vec k} \prod_j (\tr \phi^j )^{k_j}
    \, .
\]
The generalized Borchers formulation applies equally well to the
general case \eqref{invtpoly} with arbitrary $\V(\phi)$.

A complete set of physical observables for Yang-Mills theory in
any dimension are Wilson loops. These are, in particular,
interesting observables for Yang-Mills theory in $d=2$. However,
gauge invariant polynomials of the field $\phi$ form another
complete set of observables naturally suited to evaluation of the
partition function. These observables include products of $\Tr
\phi^2(x_i)$ at various points $x_i$, and more generally, traces
of any homogeneous invariant polynomial defined on the Lie algebra
$\g$. We will adopt this terminology, and refer to expectations of
products of $\Tr \phi^2(x_i)$ as \emph{$\phi$-field correlators}.
This is in marked contrast to $d=4$ Yang-Mills where the only
dimension four gauge invariant operators are $\tr (F\wedge * F)$
and $\tr (F^2)$, with the latter a topological term.

The most important property of expectation values of
gauge-invariant observables in two dimensions is that they are
almost topological. A sample calculation shows that
\be
    d \left\< \f{1}{8\pi^2} \Tr\phi^2(x) \right\>_{\eps}
    =
    \left\< \f{1}{4\pi^2}
    \Tr \phi(x) d_A\phi(x) \right\>_{\eps} = 0
    \, . \label{xind}
\ee
The action
\[
    S_{\rm top} = -\half \int  i \tr(\phi F)
\]
describes a true topological field theory whose path integral is
concentrated on flat connections $F=0$. The field $\phi$ is
sometimes denoted by $B$, in which case the Lagrangian is
proportional to $\Tr (BF)$, and the terminology \emph{BF theory}
was introduced. In the small area limit (or the $\V\to 0$ limit)
$YM_2$ reproduces the results of this topological field theory.

\subsection{The Hilbert Space of YM${}_2$} \label{ssHilSP}

We consider quantization of $YM_2$ on the cylinder with periodic
spatial coordinate of period $L$. This model is well understood
and we will make no attempt at exposition since several excellent
references exist in the literature
\cite{Blau:1993hj,Birmingham:1991ty,Cordes:1994fc,Moore:1994dk}.
Our purpose here is to point out an unexpected mathematical
relationship having to do with the space of class functions on a
Lie group that is predicted by the generalized Wightman
construction introduced earlier in the paper.

The Hilbert space of this model is known to be the space of $L^2$
class functions on $G$ with inner product
\be \lab{haar}
    \< f_1 \mid f_2 \>
    =
    \int_G dU f_1^*(U) f_2(U)
\ee
where $dU$ is the Haar measure normalized to give volume one. For
compact gauge groups, the Peter-Weyl theorem implies the
decomposition of $L^2(G)$ into unitary irreps,
\[
    L^2(G)
    =
    \oplus_{R} R \otimes \bar R
\]
Consequently a natural basis for the Hilbert space of states is
provided by the characters in the irreducible unitary
representations. This is known as the \emph{representation basis}.
The states $|R\>$ have wavefunctions $\chi_R(U)$ defined by
\be \lab{willoop}
    \< U \mid R \>  \equiv   \chi_R(U)  \equiv   \Tr_R(U)
\ee
While eqns.~\er{haar} and \er{willoop} provide two different
expressions for the inner product of YM${}_2$, a third expression
for the same inner product can be derived from the generalized
Borchers construction, in the special case of constant Wightman
functions. The sesquilinear form is given by
\eqref{sesq-form-constW}, and the inner product of YM$_2$
therefore comes from \eqref{sesq-form-constW} after taking the
quotient by zero-norm states, and subsequently, taking the
completion.

\subsection{The correlators of $YM_2$}

The correlators of $YM_2$ are determined by
representation-theoretic invariants of the gauge group such as
Casimir operators, and by the integration measure defined by the
Riemannian metric on the Riemann surface $\Sigma$. Explicit
expressions have been found by Nunes and Schnitzer
\cite{Nunes:1995pv}, using the abelianization technique for path
integrals developed by Blau and Thompson. In a particular gauge,
the two-point function for 2d $SU(N)$ Yang-Mills theory on a
Riemann surface is
\begin{eqnarray}
    \<\xi^{a}(x)\xi^{b}(y)\> = \f{e^{4}}{Z_{\Sigma_{g}}}
    \sum_{l}\dim(l)^{2-2g}
    \exp\Big({-\f{e^2}{2} A C_{2}(l)}\Big)
    \makebox[1.7in]{}
    \nonumber \\
    \times \Big[ \frac{(\rho,\rho)\delta^{ab}}{N^{2}} \,\delta^{2}_{x,y}-
    (p^{ab}(l+\rho)^{2}+m^{ab}n^2) \Big] \label{new1}
\end{eqnarray}
where $l$ is the highest weight which labels the irreducible
representation of $SU(N)$, $n$ is the total number of boxes in the
Young tableau defined by $l$, $\dim(l)$ and $C_{2}(l)$ denote
respectively the dimension and quadratic Casimir, $\rho$ is the
half-sum of the positive roots, $A$ is the area of $\Sigma$, and
\[
    p^{ab}= \Bigg\{ \begin{array}{cc}
                 \frac{-1}{N(N-1)}  & \mbox{if $a\neq b$} \\
                 \frac{1}{N}        & \mbox{if $a=b$}
                \end{array}
    \qquad m^{ab}= \Bigg\{ \begin{array}{cc}
                 \frac{1}{N(N-1)}  & \mbox{if $a\neq b$} \\
                 0                 & \mbox{if $a=b$}
                \end{array}
\]
Note that the dependence of \er{new1} on $x,y$ and on the choice
of gauge goes away after inserting the Lie algebra generators and
taking the trace.

In general, it is known that the gauge-invariant $(2p)$-point
function $\< \Tr \xi^{2p}(x) \>$ on a Riemann surface of genus $g$
takes the form
\[
    \f{e^{4p}}{Z_{\Sigma_g}} \sum_\ell \dim(\ell)^{2-2g}
    \exp\Big(-\f{e^2}{2} A C_2(\ell)\Big)
    \sum_{i=1}^p f_i(\rho) C_{2i}(\ell)
\]
where $f_i$ are rational functions of $(\rho,\rho)$.

\begin{remark} \label{remark:haar}
The physical Hilbert space of this quantum theory is a
well-defined object, and we now in principle know two ways to
calculate it. As discussed previously, the Hilbert space of this
model is known to be the space of $L^2$ class functions on $G$
with inner product \er{haar}. However, by the ideas introduced in
this paper, we expect to also recover the Hilbert space inner
product from the state on the generalized Borchers algebra that is
determined by \er{new1} and all higher-order correlators. Of
course, the physical Hilbert space is recovered only after taking
the quotient by the left-kernel of this state, which restores
gauge invariance.
\end{remark}


\section{Matrix Models and Matrix States} \label{matrixmodels}

Section \ref{subsec:matrixstates} gives the definition of matrix
states. In Sections \ref{subsec:hermitian}, \ref{subsec:dgvafa} we
recall how matrix models have come to play a prominent role in
high energy physics in the last few years. Section
\ref{subsec:mmstates} shows that, in the same sense in which
scalar quantum field theories are Wightman states, matrix models
are matrix states. Finally, Section \ref{sec:limitsmatrix} points
out that an argument due to Borchers generalizes to the
noncommutative case, showing that arbitrary states on the field
algebra, which describe non-trivial quantum field theories, are
limits of matrix states.

\subsection{Matrix States} \label{subsec:matrixstates}

\begin{definition} \label{def:MS}
Let $T \in E(\uA)$ be a state, and denote by $I(T)$ the maximal
two-sided ideal contained in $L(T)$. $T$ is called a \emph{matrix
state} if $\uA / I(T)$ is a finite dimensional algebra.
\end{definition}

The terminology comes from the fact that any finite-dimensional
$\ast$-algebra with unit is isomorphic to a sub-$\ast$-algebra of
$N \times N$ matrices for some $N$. Let $\mathfrak{h}_N$ denote
the $\ast$-algebra of $N \times N$ Hermitian matrices. We
similarly define a \emph{Hermitian} matrix state to be one for
which $\A_h / I(T) \cap \A_h$ is a finite-dimensional algebra;
this algebra will then be isomorphic to a subalgebra of
$\mathfrak{h}_N$ for some $N$.

\subsection{Hermitian Matrix Models} \label{subsec:hermitian}

A Hermitian one-matrix integral (see \cite{matrev,Kazakov:2000aq}
for a review) takes the form
\[
    Z = \int [d^{N^2} M] \exp (N \Tr S(M))
\]
where $S(M)$ is an arbitrary function. The model is said to be
\emph{solvable} if the integral can be performed explicitly, at
least in the large $N$ limit. We briefly indicate how this can be
done in the simplest case. Diagonalize $M$ via the transformation
$M = {\cal O}^+ x {\cal O}$ where $x$ is diagonal and ${\cal O}
\in U(N)$. The corresponding measure can be written as:
\[
    d^{N^2} M = d[{\cal O}]_{U(N)}\, \Delta^2(x) \prod dx_k
\]
where $\Delta(x) = \prod_{i>j}(x_i-x_j)$ is the Vandermonde
determinant. The integrand does not depend on ${\cal O}$, so
integration over ${\cal O}$ produces a group volume factor. The
remaining integral over the eigenvalues is $Z = \int[\prod_{k=1}^N
dx_k] e^{N S(x_k)} \Delta^2(x)$. In the large $N$ limit the
corresponding saddle point equation takes the form
\[
    \f{1}{N} \f{\d S}{\d x_k}
    =
    S'(x_k) + \f{1}{N} \sum_{j\ne k} \f{1}{x_k - x_j}
    = 0
\]

\subsection{Dijkgraaf-Vafa Matrix Models} \label{subsec:dgvafa}

Dijkgraaf and Vafa \cite{DV} have proposed a very simple recipe to
calculate the exact quantum effective superpotential $W(S)$ for
the glueball superfield
\[
    S = -{\tr W^{\alpha}W_{\alpha}\over 16N\pi^{2}}
\]
in the confining vacua of a large class of ${\cal N}=1$
supersymmetric Yang-Mills theories. The superpotential $W(S)$
contains highly non-trivial information about the non-perturbative
dynamics of the theory. For example, it can be used to derive
dynamical chiral symmetry breaking and calculate the tension of
the associated domain walls.

Consider the $U(N)$ or $SU(N)$ theory with one adjoint Higgs
supermultiplet $\Phi$ and a tree level superpotential of the
general form
\[
    \wt  = \sum_{p\geq 1}{g_{p}\over p} \tr\Phi^{p}=
    \sum_{p\geq 1} g_{p} u_{p}\, .
\]

Dijkgraaf and Vafa have conjectured \cite{DV} that the
superpotential $W(S)$ is the sum of zero momentum {\it planar}
diagrams of the ${\cal N}=1$ theory under consideration. In our
case, their ans\"atz for the $U(N)$ theory is a holomorphic
integral over $n\times n$ complex matrices $\phi$,
\begin{equation}
    \label{matDV}
    \exp\left( n^{2}\F /S^{2}\right) = \int_{\rm planar}
    d^{n^{2}}(\phi/\Lambda)\, \exp\Bigl[ -{n\over S}\,\wt
    (\phi,g_{p})\Bigr]\, ,
\end{equation}
from which the superpotential can be deduced,
\bel{WDV}
    W(S,\Lambda^{2},g_{p}) = -N\partial_{S}\F (S,g_{p})\, .
\ee
Here, $\Lambda$ is the complex mass scale governing the one-loop
running of the gauge coupling constant (see \cite{revN1}). For
$SU(N)$ gauge theory, the integral (\ref{matDV}) must be
restricted to traceless matrices, or equivalently one must treat
$g_{1}$ as a Lagrange multiplier.

The parameter $n$ is introduced so that the planar diagrams can be
extracted by taking the $n\rightarrow\infty$ limit. The $N$
dependence of the superpotential is then given explicitly by
(\ref{WDV}). The integral (\ref{matDV}) involves complex matrices
and couplings $g_{p}$, but the calculation is the same as for
hermitian matrices and real couplings. There is no ambiguity in
the analytic continuation because we restrict to planar diagrams.
This implies that standard matrix model techniques \cite{matrev}
do apply. A nice mathematical description of holomorphic matrix
integrals was given by Lazaroiu \cite{Lazaroiu:2003vh}.

\subsection{Matrix Models as Matrix States}
\label{subsec:mmstates}

The interesting point we wish to make in this section is that the
aforementioned matrix models (arising in string theory, condensed
matter, and other branches of physics) are special cases of the
noncommutative-target Borchers construction with a \emph{matrix
state} in the star-algebra sense, as in Definition \ref{def:MS}.
An acceptable mathematical terminology is to define the term
\emph{matrix model} to simply be the noncommutative-target
Borchers algebra with a matrix state.

Consider the noncommutative Borchers algebra $A = \uA(V,
\mathfrak{h}_N)$ into the space $\mathfrak{h}_N$ of Hermitian
matrices. Define a state $\frakw$ on $A$ by the prescription
\begin{eqnarray}
    \label{matrix-state}
    \frakw_2(a_1 \times a_2)
    &=&
    \begin{cases}
    0, & (\exists i) ~a_i \te{ is nonconstant } \\

    \sum_{I_1,I_2}  (a_1)_{I_1} (a_2)_{I_2}\, K_{I_1,I_2}, & \te{ otherwise}
    \end{cases}
    \\
     \te{ where } && \nn \\
     \label{mat-integral}
    K_{I_1,I_2} &=&
    \int_{\mathfrak{h}_N} [Da] \ a_{I_1} a_{I_2} \ e^{-S(a)}
\end{eqnarray}
where $\sum_{I_1,I_2}$ denotes a sum over all possible values of
$I_1$ and $I_2$, the measure $[Da]$ runs over $\mathfrak{h}_N$,
and $\frakw_n$ for $n > 2$ are defined analogously. Here,
$I_\alpha = (i_\alpha, j_\alpha)$ denotes a pair of indices which
together specify a matrix element. $S(a)$ denotes the action of
the matrix model, which might be $\Tr(a^2)$ in the simplest case.
We have chosen to use the notation of Section
\ref{subsec:hermitian}, which describes Hermitian matrix models;
however, the framework is completely general.

After taking the quotient by the kernel of $\frakw = (\frakw_0,
\frakw_1, \frakw_2, \ldots)$, the one-particle space is simply
$\mathfrak{h}_N$, so these are bona fide matrix states. We recover
a simple Hilbert space and operator formulation for matrix models.
The matrix model is called \emph{solvable} if the
$N^2$-dimensional integrals in \eqref{mat-integral} can be reduced
to $N$-dimensional integrals, which one then expects to evaluate
by saddle-point approximations.

\subsection{Limits of Matrix States} \label{sec:limitsmatrix}

The classic result of Borchers \cite{Borchers:cu} that matrix
states are dense in the space of all states on the Borchers
algebra generalizes to the noncommutative setting. The proof
proceeds in two steps, first showing that matrix states are dense
in the states of finite order, and then showing that the latter
are dense in the set of all states. As the results of this section
are of a topological nature, we assume throughout that $\B$ is a
normed \stalg and that the space $\A = \A(\Sigma, \B)$ introduced
in eqn.~\er{GBtestfn} is a Schwartz space of rapidly decreasing
functions. The generalized Borchers algebra is consequently
endowed with the Schwartz topology.

Given a sequence of Wightman distributions defined for fields
valued in a noncommutative space, and a state on $\B$, Section
\ref{sec:NCtarget} defines a state, which we now call $T$, on the
generalized Borchers algebra. The Hilbert space $\H_T$, field
operator $A_T$, and cyclic vector $\Om_T$ are then defined, as
usual, by the GNS construction.

\begin{definition}
$T \in E(\uA)$ is said to be of \emph{order $N$} if the family of
operators $\{A_T(f) : f \in \A\}$ contains exactly $N$ linearly
independent elements, or equivalently if
\[
    \dim\big( \A / I(T) \cap \A \big) = N
\]
\end{definition}

\begin{theorem} \label{thm:limitmatrix}
Any finite-order state $T$ on the generalized Borchers
algebra $\uA$ is a limit of matrix states.
\end{theorem}

\begin{proof}
Let $\{\H, \vp, \Om\}$ be the Hilbert space, field operator, and
cyclic vector given by the GNS-type construction described in
Section \ref{sec:NCtarget}. It is clear that if $T$ is of order
$N$, there exist $N$ distributions $t^i \in \A'$ and $N$ operators
$A^i$ on $\H$ such that
\[
    \vp(f) = \sum_{i=1}^N (t^i,f) A^i
\]
for all $f \in \A$. Let $\H_n$ be the vector space spanned by all
vectors
\[
    \{ A^{i_1} \dots A^{i_n} \Om\ :\ i_j = 1, \ldots, N, \ \
    r = 0, \ldots, n \}
\]
$\H_n$ is finite dimensional, so $\H_n$ is closed and $A^i
|_{\H_n}$ is bounded. Let $E^n : \H \to \H_n$ be the associated
orthogonal projection. Also define $\b_n$ as follows:
\[
    f \
    \xymatrix{ \ar[r]^{\b_n} & \ }
    \
    \sum_{i=1}^N (t^i, f) E^n A^i E^n
\]
Thus $\b_n$ is a continuous homomorphism of $\A$ into a
finite-dimensional matrix algebra, which is an algebra of $d(n)
\times d(n)$ matrices, where $d(n) = \dim \H_n$. We remark that
generically, $d(n)$ will be an increasing function of $n$, and the
$n \to \infty$ limit resembles a ``large $n$ limit'' of matrix
models. Define approximate states $T^{(n)}$ by
\[
    (T^{(n)}, g) = \<  \b_n(g) \>_{\Om}
\]
for any $g \in \uA$. By definition, a matrix state is an
expectation in some fixed vector of a homomorphism into a
finite-dimensional matrix algebra, so clearly $T^{(n)}$ are matrix
states. Recall that $\A_k$ was defined to be the image of
$\A^{\otimes k}$ in the universal enveloping algebra $\uA$. Note
that $T^{(n)}$ coincides with $T$ on the spaces $\A_k$ for $k \leq
n$, so $T = \lim_{n \to \infty} T^{(n)}$ and the proof is
complete.
\end{proof}

\begin{theorem} \label{thm:finiteorder}
States of finite order are dense in the set of all states on the
generalized Borchers algebra.
\end{theorem}

\begin{proof}
Let $T$ be any state, and let $\{\H, \vp, \Om\}$ be the Hilbert
space, field operator, and cyclic vector given by the GNS-type
construction described in Section \ref{sec:NCtarget}. Choose $N$
arbitrary elements $f_j \in \A$ and let $\A^N$ denote the subspace
of $\A$ generated by the $f_j$. For arbitrary $g \in \A^N$, we
have
\[
    \vp(g) = \sum_{j=1}^M F_j(g) \vp(f_j)
\]
where $\{\vp(f_j)\}_{j=1 \ldots M}$ is a minimal basis of the span
of $\{\vp(f_i)\}_{i=1 \ldots N}$ and the $F_j$ are $M$ continuous
linear functionals on $\A^N$. By the Hahn-Banach theorem, extend
each $F_j$ to a functional $t^j$ defined on all of $\A$. Define
\[
    \vp'(g) = \sum_{i=1}^M (t^i, g) \vp(f_j),
    \quad \te{ and } \quad
    T[f_1, \ldots, f_N](g) = \< \vp'(g) \>_{\Om}
\]
It is now clear that each $T[f_1, \ldots, f_N]$ is a state of
order $M$. This process defines a Cauchy net $T[\ ]$ of
finite-order states converging to $T$.
\end{proof}

If an explicit functional $S(a)$ is known so that a matrix state
$\frakw$ is given by eqns.~\er{matrix-state}-\er{mat-integral},
then we say $\frakw$ is defined by its \emph{action} $S$, and call
the associated quantum theory a \emph{matrix model}. Given a
quantum theory (possibly with noncommutative target) defined in
terms of its Wightman state, Theorems
\ref{thm:limitmatrix}--\ref{thm:finiteorder} construct a sequence
of matrix states which converge to the given state. As remarked in
the proof of Theorem \ref{thm:limitmatrix}, the $n \to \infty$
limit considered in that theorem resembles, at the level of
dimensions of the relevant algebras, a large $n$ limit of matrix
models. It would be extremely interesting if there were a generic
way to reformulate each of these matrix states as a matrix model
with action $S$. Such a formulation would allow us to choose our
favorite gauge theory, and immediately write it as a limit of
matrix models.

In fact, Kazakov in a famous paper \cite{Kazakov:2000ar} completed
precisely such a construction. Kazakov considers the
multi-component scalar field theory in four dimensions
\[
    S = N \int d^4x \ \tr((\d_\mu \phi)^2 + V(\phi))
\]
with $\phi$ a Hermitian $N \times N$ matrix-valued field, and
showed that the 4D field theory at finite $N$ is equivalent
perturbatively, graph by graph of any topology, to a one-matrix
model in the large $n$ limit. The latter may provide a
nonperturbative definition of the 4D field theory. It's far from
obvious that this is related to Theorems
\ref{thm:limitmatrix}--\ref{thm:finiteorder}, but it's possible!

\section{Conclusions}
\lab{sec:conclusions}

A classic result of mathematical physics is the reconstruction of
the Hilbert space, vacuum vector, and field operators of a quantum
field theory from a given set of distributions satisfying the
Wightman axioms \cite{Reconstruction}. In previous sections, we
have described the extension of Wightman's construction to
theories with gauge symmetry, including nonabelian pure Yang-Mills
models and matrix models, and more generally to theories with
fields or test functions valued in a noncommutative algebra.
Yang-Mills theory with an adjoint-valued Higgs field also fits
within the same framework. In every case involving quantized
Yang-Mills fields, the Lie algebra dependence of the fundamental
fields can be transferred to the test function space.

Additional difficulties are encountered in the noncommutative case
which are not present in the commutative case. The Borchers field
algebra must now be defined as a universal enveloping algebra due
to the noncommutativity of the target space, rather than a
symmetric tensor algebra. Also, a sequence of $n$-point functions
no longer suffices to define a state on the field algebra; one
must also take as part of the data of the theory a state on the
target space $\B$. For usual Yang-Mills theories, this additional
target state is the trace in the adjoint representation of the
gauge symmetry algebra.

This construction provides a unified algebraic framework for
formulating properties of a broad class of quantum field theories.
Yang-Mills theory (including matrix models) and constructive field
theory models \cite{GJ} possess the following common structure:

\begin{enumerate}
\item A $\ast$-algebra $\B$ (not necessarily commutative), with the
generalized Borchers algebra $\uA$ of functions into $\B$.

\item A functional $\omega \in \uA'$, which is defined in terms of a sequence of
$n$-point functions and which satisfies $\omega(f^* \times f) \geq
0 \ (\forall f)$ in theories with no gauge symmetry, or in gauge
theories expected to possess a positive-definite inner product.
For nonpositive models such as Gupta-Bleuler QED, the functional
$\omega$ is postulated to satisfy axioms described in Section
\ref{sec:hssc}.

\item
A symmetry group $\SymmetryGp$ and a representation $\alpha :
\SymmetryGp \to \Aut(\uA)$, which is \emph{$\omega$-invariant} in
the sense that $\omega(\alpha_g(f)) = \omega(f)$ for all $f \in
\uA$ and $g \in \SymmetryGp$.

\item  A collection of ideals $I_1, \ldots, I_n$ of the algebra $\uA$
which are required to lie in the kernel of $\omega$. (Each ideal
represents a physical property satisfied by the $n$-point
functions; in Wightman QFT, $n = 2$ and the two ideals represent
locality and the positive light-cone spectral condition.)

\item A Hilbert space $\Hphys$ defined to be the completion of
$\uA / L(\omega)$, with inner product $\omega(f^* \times g)$ and
vacuum vector $\Om$.

\item Field operators defined by the GNS construction, with vacuum
expectation values equal to the Wightman functions used to define
the state $\omega$.
\end{enumerate}

The discussion following Theorem \ref{thm:finiteorder} outlines a
possible new research direction, concerned with the question of
how quantum field theories can be written as limits of matrix
models. It would be very interesting to have a deeper
understanding of the issue raised by Remark \ref{remark:haar}.
Moreover, this construction could be studied for fields which are
sections of nontrivial principle fibre bundles, as suggested by
Remark \ref{remark:vectorbundle}.

We anticipate possible connections between this work and
deformation theory, which we elaborate on briefly here. If we
consider continuous deformations of the product $\cdot_{\B}$ in
the noncommutative target space $\B$, with respect to a parameter
$\ve$, then the constructions in this paper define continuous
families of Wightman functionals $\w_\ve$ and associated quantum
field theories with Hilbert space inner products $\<\ ,\ \>_\ve$.
Consider the simple deformation which multiplies the associative
algebra product (and hence the Lie bracket) in $\B$ by $\ve$. For
each $\ve$, the Poincar\'e-Birkhoff-Witt theorem implies an
isomorphism $\uA_\ve \simeq S\A$, where $S\A$ is the symmetric
tensor algebra over $\A$. Multiplication on $\uA_\ve$ defines a
family of multiplications on $S\A$,
\bel{def}
    f \ast_\ve g = fg + \f12 \ve \{ f, g\} + \sum_{k\geq 2} \ve^k
    B_k (f,g)
\ee
for some bilinear forms $B_k$. This well-known construction is
called deformation quantization in the direction of the Poisson
bracket.

In general, given a quantum field theory into a noncommutative
space $\B$, one could consider deforming the multiplication on the
target space, and in certain cases, we expect that the $\eps$
dependence might factor out of the Wightman functions and the
Hilbert space inner product will scale in some simple way with
respect to the deformation parameter $\eps$.

\section*{Acknowledgements}

The author wishes to thank his graduate advisor, Arthur Jaffe, for
continuous support and encouragement, and to thank Daniel Jafferis
for many useful discussions, especially on matrix models.

\def\pr#1#2#3{{\it Phys.\ Rep.\ }{\bf #1} (#2) #3}
\def\npps#1#2#3{{\it Nucl.\ Phys.\ Proc.\ Suppl.\ }{#1} (#2) #3}

\end{document}